# Electroluminescence and Photocurrent Generation from Atomically Sharp WSe₂/MoS₂ Heterojunction *p-n* Diodes

Rui Cheng[1], Dehui Li[2], Hailong Zhou[2], Chen Wang[1], Anxiang Yin[2], Shan Jiang[2], Yuan Liu[1], Yu Chen[1], Yu Huang[1,3], Xiangfeng Duan[2,3,*]

*[1]Department of Materials Science and Engineering, University of California, Los Angeles, CA 90095, USA; [2]Department of Chemistry and Biochemistry, University of California, Los Angeles, CA 90095, USA; [3]California Nanosystems Institute, University of California, Los Angeles, CA 90095, USA*

*Corresponding email: xduan@chem.ucla.edu

**TOC Figure**

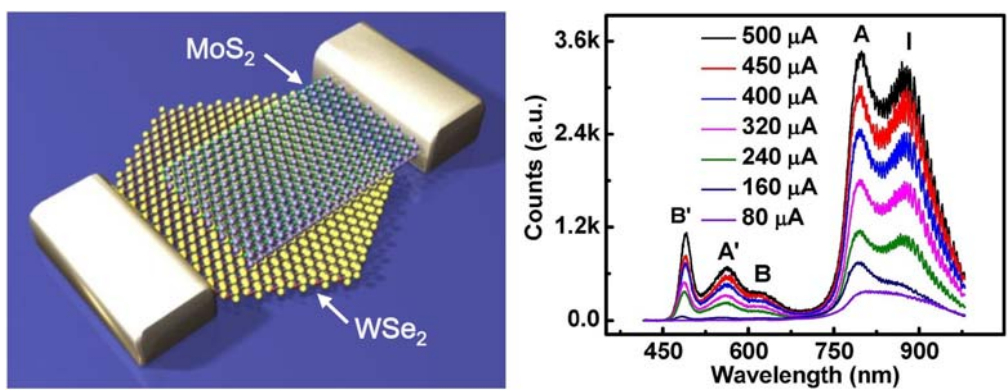

**Abstract.** The *p-n* diodes represent the most fundamental device building block for diverse optoelectronic functions, but are difficult to achieve in atomically thin transition metal dichalcogenides (TMDs) due to the challenges in selectively doping them into *p-* or *n-*type semiconductors. Here we demonstrate that an atomically thin and sharp heterojunction *p-n* diode can be created by vertically stacking *p*-type monolayer tungsten diselenide (WSe₂) and *n*-type few-layer molybdenum disulfide ($MoS_2$). Electrical measurements of the vertically staked $WSe_2$/$MoS_2$ heterojunctions reveal excellent current rectification behaviour with an ideality factor of 1.2. Photocurrent mapping shows rapid photoresponse over the entire overlapping region with a highest external quantum efficiency up to 12%. Electroluminescence studies show prominent band edge excitonic emission and strikingly enhanced hot electron luminescence. A systematic investigation shows distinct layer-number dependent emission characteristics and reveals important insight about the origin of hot-electron luminescence and the nature of electron-orbital interaction in TMDs. We believe that these atomically thin heterojunction p-n diodes represent an interesting system for probing the fundamental electro-optical properties in TMDs, and can open up a new pathway to novel optoelectronic devices such as atomically thin photodetectors, photovoltaics, as well as spin- and valley-polarized light emitting diodes, on-chip lasers.

**Keywords:** $WSe_2$ $MoS_2$; heterojunction; electroluminescence; photocurrent; Van der waals

Two-dimensional layered materials, such as graphene, $MoS_2$, and $WSe_2$, are emerging as an exciting material system for a new generation of atomically thin optoelectronics, including photodetectors,[1-7] ultrafast lasers,[8] polarizers,[9] touch panels[10] and optical modulators[11] due to their unique electronic and optical properties.[12-23] In this regard, the monolayer transition metal dichalcogenides (ML-TMDs) is particularly interesting due to their direct energy bandgap and the non-centrosymmetric lattice structure.[12-13] The *p-n* diodes represent the most fundamental device building blocks for most optoelectronic functions, including photodiodes and light emitting diodes. However, it is particularly difficult to create *p-n* diodes in atomically thin TMDs due to the challenges in selectively doping them into p- or n-type semiconductors. Contact engineering has been explored to created *p-n* diodes in TMD based layered materials.[24] Electroluminescence (EL) from ML-MoS$_2$ has been reported in a metal-MoS$_2$ Schottky junction through a hot carrier process.[25] Electrostatic doping has also been used to create planar *p-n* diodes, but usually with relatively gradual doping profile (limited by the fringe electrical field) and typically relatively low optoelectronic efficiency (e.g. photon to electron conversion external quantum efficiency (EQE) ~0.1–1%).[26-28]

The atomically thin geometry of these 2D materials can allow band structure modulation in a vertically stacked heterostructures to form atomically sharp heterojunctions.[29] For example, this strategy allows gapless graphene to be used in field-effect tunnelling devices,[29,30] barristors,[31] inverters,[32] and photodetectors[7] while staked together with other 2D materials in the vertical direction. Although the nearly perfect 2D structure and low density of states in graphene provide advantages in some heterostructure devices, its gapless nature prevents the formation of a large potential barrier for charge separation and current rectification. On the other hand, the vertical heterojunction *p-n* diode formed between a TMD material and a bulk material has recently been reported, but usually with no EL[33,34] or very weak EL.[35]

Here we report an atomically thin *p-n* diode based on a heterojunction between synthetic *p-*type ML-WSe$_2$ and exfoliated *n-*type MoS$_2$ flake. The atomically thin *p-n* diode exhibit well defined current rectification behaviour and can enable efficient photocurrent generation with an EQE up to 12%. Unlike the planar structures where the active area is confined to the lateral interface region, photocurrent mapping shows fast photoresponse and demonstrates that the *p-n* junction is created throughout the entire $WSe_2/MoS_2$ overlapping area. Furthermore, prominent EL is observed with rich spectral features that can reveal important insights about electron-orbital interaction in TMD based materials.

The vertical heterojunction *p-n* diode is formed between synthetic *p*-type ML-$WSe_2$ and exfoliated *n*-type $MoS_2$ flake (Figure 1a,b). Triangular domains of ML-$WSe_2$ was first synthesized on 300 nm Si/$SiO_2$ substrate typically with a bilayer (BL) region in the centre (see Figure 2a), which were characterized by using optical microscope, atomic force microscopy (AFM), and Raman spectroscopy (Figure S1). Mechanically exfoliated $MoS_2$ flakes were then transferred onto synthetic $WSe_2$ domains to form vertically stacked heterojunctions. Electron-beam lithography and electron beam evaporation were used to define the contact electrodes. A thin Ni/Au film (5nm/50 nm) and Au film (50nm) were used as the electrode for $MoS_2$ flake[36] and $WSe_2$ domain to form Ohmic contacts with minimized contact resistance and potential barrier (Figure 1b). Figure 1c shows the ideal band diagrams of the heterojunction *p-n* diode at zero bias. The built-in potential and applied voltage are mainly supported by a depletion layer with abrupt atomic boundaries, and outside the boundaries the semiconductor is assumed to be neutral.

Figure 2a shows an optical microscopy image of a synthetic $WSe_2$ domain on 300nm Si/$SiO_2$ substrate. A triangular shaped BL-$WSe_2$ domain was typically observed in the centre of the triangular ML-$WSe_2$ domain, indicating the nearly perfect lattice structure of the synthetic $WSe_2$. Figure 2b shows a top-view scanning electron microscopy (SEM) image of the vertical heterojunction. The $MoS_2$, $WSe_2$ layers and the contact electrodes are labelled with different artificial colours to highlight the device structure.

Photoluminescence (PL) mapping was used to illustrate the stacking structure of $WSe_2/MoS_2$ heterojunction (Figure 2c). The PL mapping shows distinct PL emission from $WSe_2$ (red region in Figure 2c) and $MoS_2$ region (green region in Figure 2c), with a consistent structure layout as that observed in the SEM image (Figure 2b). The uniform PL from $WSe_2$ and $MoS_2$ also indicates the excellent quality of the TMD materials. The PL spectra of $WSe_2$ show a strong layer-number dependence (Figure 2d), with the PL intensity in ML-$WSe_2$ at least 10 times stronger than that in BL-$WSe_2$. The PL spectrum in ML-$WSe_2$ shows a peak at 785 nm, corresponding to the "A" exciton peak.[37] The PL in BL-$WSe_2$ also exhibits the "A" exciton peak with an additional broad peak at ~877 nm, which is attributed to indirect band gap emission (typically label as "I" peak).[16] These PL studies are consistent with previous experimental studies and theoretical calculations,[16,37] indicating the good crystalline quality of the synthetic $WSe_2$. The PL spectrum from $MoS_2$ flake shows a peak at 677 nm, corresponding to "A" exciton peak in $MoS_2$. It is also important to note that the "B" exciton peak can be observed in both $WSe_2$ (605 nm) and $MoS_2$ (620 nm), but with the intensity typically 2-3 order magnitude weaker than the "A" exciton peak (Figure. 2d inset).

We have further characterized the atomic structure of the stacked heterojunction using cross-sectional transmission electron microscope (TEM) studies. The high resolution TEM image clearly shows the $WSe_2/MoS_2$ heterojunction with a 13-layer $MoS_2$ flake on top of BL-$WSe_2$ (Figure 2e). Energy dispersive X-ray spectroscopy (EDS) was further used to analyze the elemental distribution across the heterojunction interface. An EDS elemental line scan in vertical direction shows a rather narrow tungsten distribution profile located near the heterojunction interface (Figure. 2f), with the full width at half maximum ~1.2 nm, corresponding to BL-WSe$_2$ observed in Figure 2e. Together, these structural and PL characterisations demonstrate that the atomically sharp heterojunctions are formed by vertically stacking atomically thin WSe$_2$ and MoS$_2$.

Before testing the electrical characteristics of the heterojunction *p-n* diodes, we have first characterized the electrical transport properties of MoS$_2$ and WSe$_2$ to ensure Ohmic contacts were achieved. To this end, the MoS$_2$ and WSe$_2$ field effect transistors (FETs) were fabricated on Si/SiO$_2$ substrate, with Ni/Au thin film as the source-drain contacts for MoS$_2$, and Au thin film as the contacts for WSe$_2$, and the silicon substrate as a back gate electrodes. Figure 3a and 3b show the $I_{ds}$-$V_{ds}$ characteristics at varying back gate voltages for MoS$_2$ and WSe$_2$, respectively. Importantly, a linear $I_{ds}$-$V_{ds}$ relationship is clearly observed for both MoS$_2$ and WSe$_2$, indicating Ohmic contacts are achieved for both materials. The formation of Ohmic contacts for both MoS$_2$ and WSe$_2$ is very important, since the Schottky barrier at the contact area may severely affect the electronic and optoelectronic characteristics of the vertical heterojunction and could induce photocurrent generation or EL at the contact region.[25] Furthermore, $I_{ds}$-$V_{ds}$ plots at varying back gate voltage show that the current increases with increasing positive gate voltage for MoS$_2$, indicating an *n*-type semiconductor behaviour. On the contrary, the current increases with decreasing negative gate voltage in WSe$_2$ FET, consistent with the *p*-type characteristics.

With Ohmic contacts formed for both MoS$_2$ and WSe$_2$, we continue to probe the electrical transport properties of the heterojunction *p-n* diode. Importantly, a clear current rectification behaviour is observed in ($I_{ds}$-$V_{ds}$) plots for the WSe$_2$/MoS$_2$ heterojunction (Figure 3c), with current only being able to pass through the device when the *p*-type WSe$_2$ is positively biased. The observation of current rectification clearly demonstrates a *p-n* diode is formed within the atomically thin WSe$_2$/MoS$_2$ heterojunction. The ultrathin nature of the heterojunction allows gate tunability of the diode characteristics. The diode output characteristic ($I_{ds}$-$V_{ds}$) under different back gate voltage show that the output current decreases with increasing positive gate voltage, suggesting that the *p*-type WSe$_2$ is partly limiting the charge transport in the device.

The $I_{ds}$-$V_{ds}$ output characteristics of the vertical heterojunction under forward bias can be viewed as a vertical heterojunction *p-n* diode in series with an additional *p*-type FET due to the side contact on WSe$_2$. In general, the heterojunction *p-n* diode resistance decreases exponentially with increasing bias voltage, and the series p-FET resistance is nearly constant with bias voltage. Therefore, the heterojunction resistance is dominated by *p-n* diode at low bias and dominated by the *p*-type WSe$_2$ FET under high forward bias. We have also fitted the diode characteristics and calculated the ideality factor of our heterojunction device based on the model of a *p-n* diode with a series resistor (Figure. 3d). Importantly, an ideality factor of n=1.2 was derived with a series resistance of 80 M$\Omega$, at zero gate voltage, and an ideality factor of n=1.3 was derived with a series resistance of 33 M$\Omega$, at -20V gate voltage. The achievement of ideality factor close to 1 indicates the excellent diode behaviour of our atomically sharp heterojunction *p-n* diode. The decrease of the series resistance with increasing negative gate voltage is also consistent with our model that the *p*-type WSe$_2$ is the limiting series resistor at high forward bias.

The electrical measurements indicate excellent diode behaviour in the atomically thin vertical heterojunction. To further characterize the diode characteristics in our vertical heterojunction, the photocurrent mapping was carried out at zero bias under a confocal microscope. Figure 4a shows an optical microscope image of the $WSe_2$/$MoS_2$ heterojunction depicting the relative position between $WSe_2$, $MoS_2$ and the electrodes. The corresponding photocurrent mapping at zero bias with a 514 nm laser excitation (5 μW) is shown in Figure 4b, with the ML-$WSe_2$ region outlined by purple dotted line, few-layer $MoS_2$ outlined by blue dotted line, and the electrodes outlined by golden solid lines. The photocurrent mapping shows clear photoresponse from the entire overlapping region, indicating the formation of a broad area *p-n* junction across the entire $WSe_2$/$MoS_2$ overlapping area. It is also interesting to note that the photocurrent in ML-$WSe_2$/$MoS_2$ region is much stronger than that in BL-$WSe_2$/$MoS_2$ region, suggesting that the direct band gap plays an important role in the photocurrent generation process.[6,7] A detailed understanding of the different response of ML vs. BL-$WSe_2$/$MoS_2$ will be an interesting topic for future studies. No measurable photocurrent was observed from the non-overlapping regions (only $WSe_2$ or $MoS_2$) or the electrical contacts, which is expected for zero bias photocurrent since the photogenerated carries in the regions outside *p-n* junction cannot be effectively separated and extracted.

The output characteristics ($I_{ds}$-$V_{ds}$) of the vertical heterojunction with and without laser illumination (514 nm, 5 μW) show clear photovoltaic effect with an open-circuit voltage of ~0.27 V and a short-circuit current of ~ 0.22 μA (Figure. 4c). In general, the photoresponse exhibits a rapid temporal response beyond our experimental time resolution of 100 µs (Figure. 4c inset), demonstrating that the photoresponse is originated from photocarrier generation rather than any other extrinsic effects. Based on the photocurrent response and input laser power, we can determine the external quantum efficiency (EQE) of the photon to electron conversion. The EQE ($\eta$) is defined as the ratio of the number of carriers collected by electrodes to the number of the incident photon, or

$\eta=(I_{ph}/q)/(P/h\nu)\times100\%$ where $I_{ph}$ is the photocurrent, h is Planck's constant, $\nu$ is the frequency of light, q is the electron charge and P is the incident light power. Our study showed that the EQE in our vertical heterojunction can reach 11% under a 514 nm laser excitation with power of 5 µW. Furthermore, it is found that the EQE increases with decreasing excitation laser power and decreases with increasing excitation power (Figure. 4d), with a maximum EQE of 12% observed under an excitation power of 0.5 µW. The decreasing EQE with increasing excitation power could be attributed partly to absorption saturation in $WSe_2$ and partly to the screening of the built-in electric field by the excited holes in the valence band of $WSe_2$.[38] The power dependent EQE of the same device under 633 nm excitation shows a similar trend but with generally lower values than those under 514 nm excitation, which may be attributed to the spectral dependent optical absorption coefficient.[6] It is important to note that the EQE observed in the vertical $WSe_2/MoS_2$ heterostructure devices is much higher than those in lateral electrostatically doped $WSe_2$ *p-n* homojunctions (0.1-3%),[27,28] which may be partly attributed to more efficient charge separation resulting from an atomically sharp vertical *p-n* junction. In contrast, the electrostatic doping would typically exhibit a spatial doping gradient, and is difficult to achieve atomically sharp junctions.

The above electrical transport and photocurrent studies demonstrate excellent *p-n* diode characteristics in the atomically sharp $WSe_2/MoS_2$ heterojunction. Since *p-n* diode represents the basic device element for a light-emitting diode, we have further investigated the electroluminescence from these heterojunction *p-n* diodes. Figure 5a shows an EL image acquired under a forward bias of 3V and a forward current of ~100 μA. The shapes of $WSe_2$, $MoS_2$ and gold electrodes were outlined in the same way as before to identify the position of the EL. In contrast to the photocurrent generation from the entire overlapping area, it is important to note that the EL is localized at the overlapping area in close proximity to the electrodes. This can be explained by the electric field distribution in the heterojunction under different bias. For photocurrent mapping at zero bias (or a small bias less than the turn on voltage), the *p-n* diode junction resistance dominates the entire device, and therefore photocurrent can be seen from the entire overlapping area where there is a *p-n* junction. For EL studies at much higher forward bias exceeding the *p-n* diode turn-on voltage, the resistance of the ML-$WSe_2$ becomes an increasingly important component of the total resistance. Therefore, the most voltage drop occurs across the heterojunction edge near the electrodes due to the large series resistance of the ML-$WSe_2$. This is also consistent with the result recently reported for $MoS_2/Si$ heterojunctions.[35]

Figures 5b and 5c show the EL spectra of a ML- and a BL-$WSe_2/MoS_2$ heterojunction with increasing injection current. The plot of the overall EL intensity as a function of injection current shows an apparent threshold (Figure. 5d), with little EL below the threshold, and linear increase above threshold. The threshold current may be explained by the band alignment of the heterojunction under different bias voltages (Figure. 5e,f). In general, due to different band gap and band alignment among the conduction band and valence band edge, the barrier for hole transport across the junction is smaller than that for the electrons. With increasing forward bias (below a certain threshold), the holes from $WSe_2$ are first injected into $n$-type $MoS_2$ region, while few electrons can overcome the barrier to reach $WSe_2$ (Figure. 5e). Due to the nature of indirect band gap in few-layer $MoS_2$, the yield of radiative recombination is relatively low at this point. As a result, the EL intensity is very low when the hole injection dominates the charge transfer across the heterojunction. With further increasing bias across the heterojunction (above electron injection threshold), the conduction band of $MoS_2$ is shifted upward, both electrons and holes can go cross the heterojunction and are injected into $p$-type and $n$-type region respectively (Figure. 5f). At this point, the radiative recombination in $WSe_2$ dominates the EL with its intensity increasing linearly with the injection current. It is noted that the EL intensity observed in $ML$-$WSe_2/MoS_2$ heterojunction is much stronger than that in $BL$-$WSe_2/MoS_2$ heterojunction due to the higher radiative recombination rate in direct band gap $ML$-$WSe_2$ vs indirect bandgap $BL$-$WSe_2$.

The EL spectra show rich spectral features and can be well fitted using multiple Guassian functions (Figure S2) with five main peaks, which can be assigned as excitonic peaks A (~792 nm) and B (~626 nm), hot electron luminescence (HEL) peaks A' (~546 nm) and B' (~483 nm) and an indirect band gap emission peak I (~880 nm). The A exciton peak dominates the spectra of the EL in $ML$-$WSe_2$ (Figure 5a) while the indirect band gap emission I is significant in $BL$-$WSe_2$ (Figure 5b). Strikingly, the EL spectra of both $ML$- and $BL$-$WSe_2$ show prominent B exciton peak and HEL A', B' peaks, which are usually 100-1000 times weaker than the A exciton peak in the PL measurements[37] and have not been reported in EL previously. In contrast, the EL spectra of the vertical heterojunction show that the intensities of these HEL peaks are only 3-10 times weaker than the A exciton peak, suggesting a relative enhancement of the HEL by about two orders of magnitude in our EL studies, which may be attributed to the electric field induced carrier redistribution.[39]

The origin of the HEL peaks in TMD materials remains the subject of debate and is difficult to probe due their low emission probability.[37, 40-42] The HEL peaks A' and B' are generally believed to arise from the splitting of the ground and excited states of A and B transitions due to the electron-orbital interaction via either inter- or intralayer perturbation or both.[40,41] However, there is no yet clear evidence to prove which perturbation dominates the electron-orbital interaction. The emergence of intense HEL emission in our ML-$WSe_2$/$MoS_2$ and BL-$WSe_2$/$MoS_2$ heterojunction can offer a new platform to probe the origin of HEL peaks and the nature of electron-orbital interaction in TMDs. The presence of HEL peaks A' and B' in EL spectra of ML-$WSe_2$/$MoS_2$ heterojunction (Figure 5b) indicates that intralayer perturbation plays a role in the formation of these HEL peaks. On the other hand, it is noted that the relative intensities of HEL peaks (comparing with the respective A peak) in BL-$WSe_2$/$MoS_2$ heterojunction (Figure 5c) are clearly much stronger than that in ML-$WSe_2$/$MoS_2$ heterojunction, suggesting that interlayer perturbation may also contribute to the HEL peaks (which can be further supported by temperature dependent studies, see below).

To further probe physical mechanism governing the photon emission process in the atomically thin *p-n* diode, we have also conducted the temperature dependent EL studies at 25, 50 and 75 °C for both the ML- and BL-WSe$_2$/MoS$_2$ heterojunctions (Figure 5g, h), and plotted the normalized peak intensities for A and B' peaks as a function of temperature (Figure 5i). For ML-WSe$_2$/MoS$_2$ heterojunction, the EL intensity of all spectral peaks show a consistent decrease with increasing temperature (Figure 5g,i), which is a common phenomenon in the LED devices and can be attributed to the exponential enhancement of nonradiative recombination rate with increasing temperature.[39] In striking contrast, temperature dependent EL in the BL-WSe$_2$/MoS$_2$ heterojunction displays highly distinct features. First, the A exciton peak in BL-WSe$_2$/MoS$_2$ heterojunction shows an unusual increase (instead of decrease) with increasing temperature. (Figure 5h,i). This increase in A exciton emission may be explained by thermally decoupling neighboring layers via interlayer thermal expansion, which can induce a band gap crossover from the indirect gap to the direct one with the increasing decoupling at higher temperature. A similar thermal decoupling effect has been observed in MoSe$_2$ by PL studies.[43] Second, the HEL peak B' (and A') shows a much greater decrease with increasing temperature than that in ML -WSe$_2$/MoS$_2$, indicating the weakening electron-orbital interaction with the decoupling neighboring layers. These temperature dependent characteristics are consistent seen in three devices studies and further suggests that that the interlayer perturbation plays an important role in electron-orbital interaction in WSe$_2$, which is consistent with the observation of strong interlayer excitons in TMD based heterojunctions.[44,45]

In summary, we have fabricated WSe$_2$/MoS$_2$ heterojunction *p-n* diodes with atomically thin geometry and atomically sharp interface. The scanning photocurrent measurement demonstrates that the *p-n* junction was formed over the entire overlapping area with a maximum photon-to-electron conversion EQE of 12%. The EL measurement allows for the identification of emission from different optical transitions. Hot electron luminescence peaks were observed in EL spectra of $WSe_2$ for the first time and used to investigate the electron-orbital interaction in $WSe_2$. Our novel heterojunction structure offers an interesting platform for fundamental investigation of the microscopic nature of the carrier generation, recombination and electro-optical properties of single or few-layer TMD materials, and can open up a new pathway to novel optoelectronic devices including atomically thin photodetectors, photovoltaics, as well as spin- or valley-polarized light emitting diodes and on-chip lasers.

Note added: During the finalization of this manuscript we became aware of two related studies.[46-47]

**Methods**

**Fabrication of the vertical heterostructure devices.** To fabricate the vertical $WSe_2/MoS_2$ heterojunction devices, $WSe_2$ was grown using a physical vapor deposition process on a $Si/SiO_2$ (300 nm $SiO_2$) substrate. 0.2 g $WSe_2$ powder (Alfa Aesar, 13084) was added into an alumina boat as precursor. The blank $Si/SiO_2$ substrates (1 cm by 5 cm) were loaded into a home-built vapor deposition system in a horizontal tube furnace (Lindberg/Blue M) with 1-inch quartz tube. The system was pumped down to a vacuum of 10 mTorr in 10 min, and re-filled with 150 sccm of ultra-high purity argon gas (Airgas, ~ 99.9999%) then heated to desired growth temperature within 30 min. After that, the growth kept at the desired temperature for 30 min, and then terminated by shutting off the power of the furnace. The sample was naturally cooled down to ambient temperature. The $MoS_2$ flakes were then exfoliated onto the WSe$_2$ flakes through a micromechanical cleavage approach. The metal electrodes (for probe contact or wire bonding purposes) were patterned on the Si/SiO$_2$ substrates by using electron-beam lithography and electron-beam deposition of Ti/Au (50/50 nm) thin film. Ni/Au (5/50 nm) contact electrodes were then deposited to form the Ohmic contact to MoS$_2$ and Au (50 nm) was deposited to form the Ohmic contact to WSe$_2$.

**Microscopic, electrical, optical, and optoelectrical characterizations.** The microstructures and morphologies of the nanostructures are characterized by a JEOL 6700 scanning electron microscope (SEM). The cross-section image of the heterostructure device is obtained by an FEI Titan transmission electron microscope (TEM). The D.C. electrical transport measurements were conducted with a Lakeshore probe station (Model TTP4) and a computer-controlled analogue-to-digital converter (National Instruments model 6030E). The confocal micro-PL and Raman measurements were conducted on a Horiba LABHR confocal Raman system with 600g/mm grating, 50× diffraction-limited objective (N.A.=0.75), with an Ar laser (514 nm) or a He-Ne laser (633 nm) excitation. The scanning photocurrent measurements were conducted with the same Horiba LABHR confocal Raman system combined with the same electrical measurement system. The EL measurements were performed on a home-build confocal PL measurement system combining with the same electrical measurement system with a temperature control in Ar environment. Unless mentioned in the main text, all measurements were conducted at room temperature. The EL images were collected by a 50× objective (N.A.=0.5) and captured by a liquid Nitrogen cooled CCD camera (Princeton instruments PyLoN 400F). The spectra were taken by using an Acton 2300i spectrometer with 150g/mm grating and liquid Nitrogen cooled CCD.

**Supporting Information Available:** The AFM and Raman characterizations of synthetic

WSe$_2$ domain on 300 nm Si/SiO$_2$ substrate and the analysis and peak fittings of the EL

spectra. This material is available free of charge *via* the Internet at http://pubs.acs.org.

**Figures & Legends**

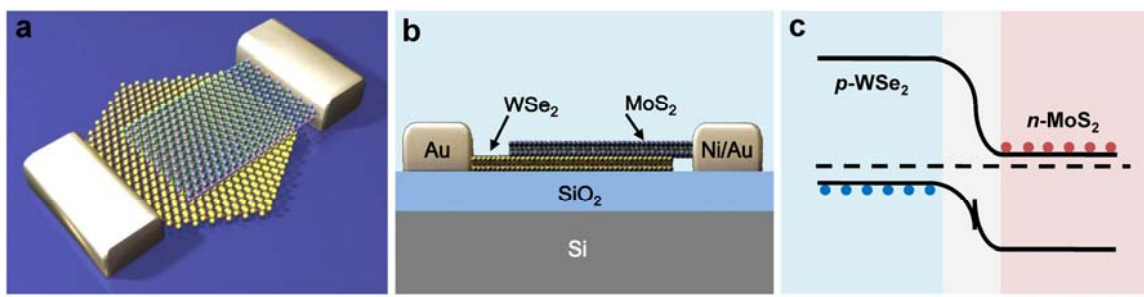

Figure 1. Schematic illustration and band diagram of the WSe$_2$/MoS$_2$ vertical heterojunction *p-n* diode. (a) A schematic illustration of the WSe$_2$/MoS$_2$ vertical heterojunction device show that a transferred MoS$_2$ flake on synthetic WSe$_2$ forms a vertical heterojunction. (b) A schematic illustration the cross-sectional view of the WSe$_2$/MoS$_2$ vertical heterojunction device. (c) The ideal band diagram of WSe$_2$/MoS$_2$ heterojunction *p-n* diode under zero bias.

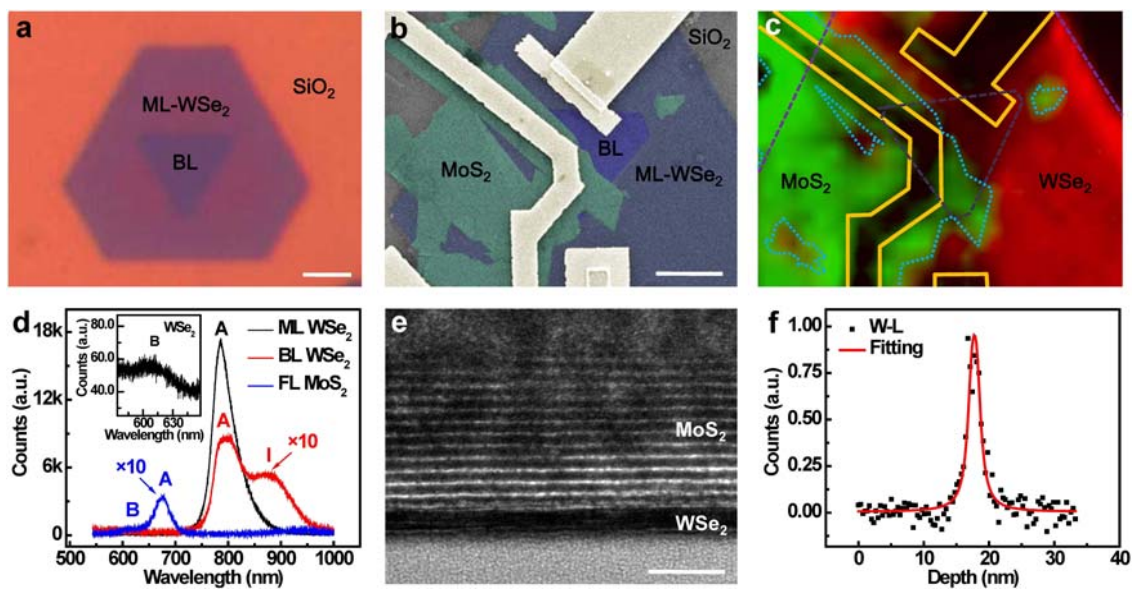

Figure 2. Structural characterization of the WSe₂/MoS₂ heterojunction *p-n* diode. (a) Optical microscopy image of a truncated triangular domain of monolayer WSe₂ with an inverted triangular bi-layer region at the centre. (b) The false colour SEM image of the WSe₂/MoS₂ vertical heterojunction device, with ML-WSe₂ highlighted by blue colour, BL-WSe₂ area by violet colour, MoS₂ by green colour, and metal electrodes by golden colour. The scale bar is 3 μm. (c) The PL mapping of the WSe₂/MoS₂ heterojunction device, with red colour representing the PL from MoS₂ and the green colour representing PL from WSe₂. (d) The PL spectra of synthetic ML- and BL-WSe₂ and few-layer MoS₂ flakes with the A, B exciton peaks and indirect transition I peak labelled. The intensities of BL-WSe2 and FL-MoS2 are multiplied by 10 times for better visibility. Inset, the B exciton peak in ML- WSe₂. (e) The cross-sectional HRTEM image of the WSe₂/MoS₂ heterojunction interface. The scale bar is 5 nm. (f) The EDS element distribution profile from the bottom to the top of Fig. 2e. The black square represent the distribution profile of W-L characteristic peaks. The red line represents the fitting curve for W-L distribution profile, with a full width at half maximum of 1.2 nm, corresponding to bilayer WSe₂.

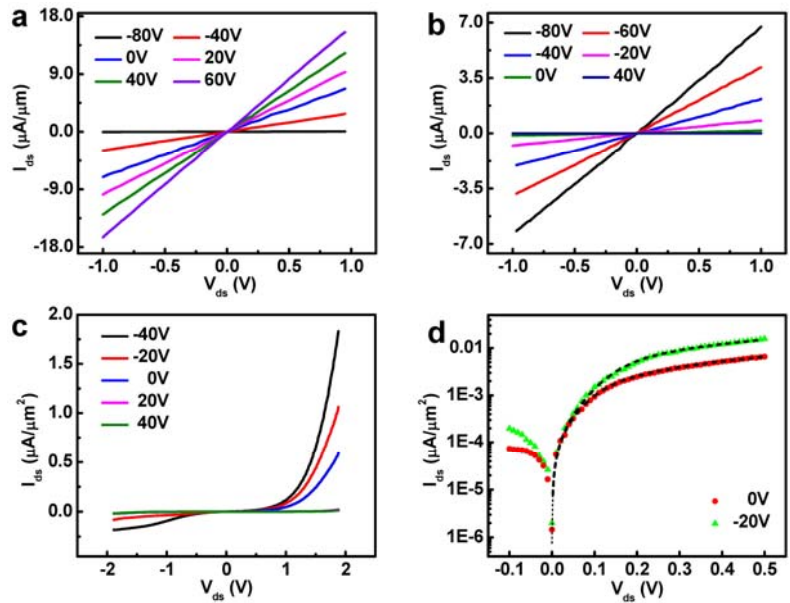

Figure 3. Electrical characterization of the WSe$_2$/MoS$_2$ heterojunction *p-n* diode. (a) The $I_{ds}$-$V_{ds}$ characteristics of *n*-type MoS$_2$ FET transistor with Ni/Au (5/50 nm) contacts. (b) The $I_{ds}$-$V_{ds}$ characteristics of *p*-type WSe$_2$ FET transistor with Au (50 nm) contacts. (c) Gate-tunable output characteristics of the WSe$_2$/MoS$_2$ heterojunction *p-n* diode. (d) The derivation of the *p-n* diode ideality factor by using a model consists of an ideal *p-n* diode with a series resistor. An ideality factor of 1.2 was derived with a series resistor of 80 MΩ at 0 V gate voltage (red circle), and an ideality factor of 1.3 was derived with a series resistor of 33 MΩ at -20 V gate voltage (green triangle).

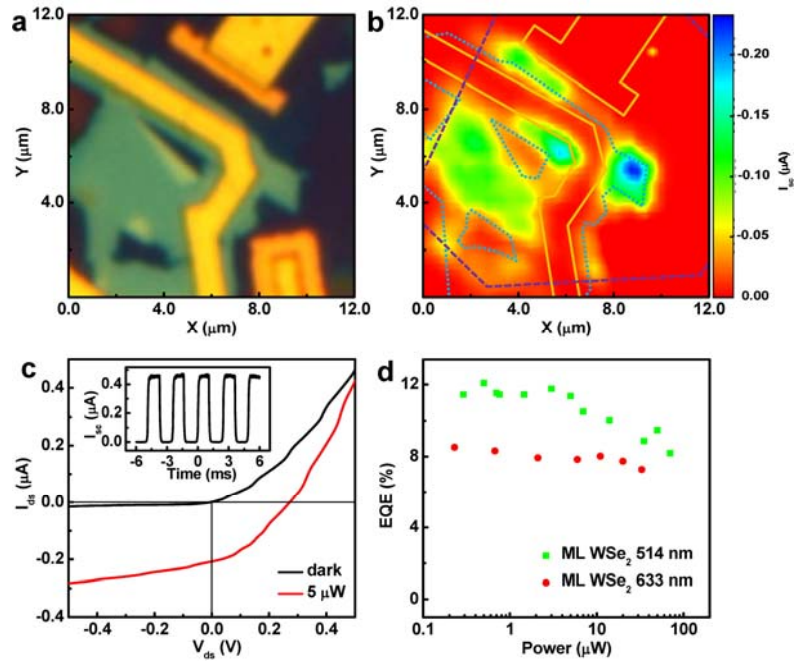

Figure 4. Photoresponse of the WSe₂/MoS₂ heterojunction *p-n* diode. (a) Optical microscpope image of the WSe₂/MoS₂ heterojunction. (b) False colour scanning photocurrent micrograph of the WSe₂/MoS₂ heterojunction device acquired at $V_{ds}$=0 V and $V_{BG}$=0 V under irradiation 514 nm laser (5 μW). The purple square dotted line outlines the ML-WSe₂ and the dark purple square dotted line outlines the BL-WSe₂. The blue circle dotted line outlines the MoS₂ and the golden solid line outlines the gold electrodes. Photocurrent were observed in the entire overlapping junction area. (c) Experimental output ($I_{ds}$-$V_{ds}$) characteristic of the vertical heterojunction device in the dark (black) and under illumination (wavelength: 514 nm; power, 5 μW). Inset, temporal response of the photocurrent generation under 514 nm illumination (10μW). (d) Power-dependent EQE of the heterojunction device under 514 nm and 633 nm laser excitation wavelengths at $V_{ds}$=0 V and $V_{BG}$=0 V. A maximum EQE of 12% was observed.

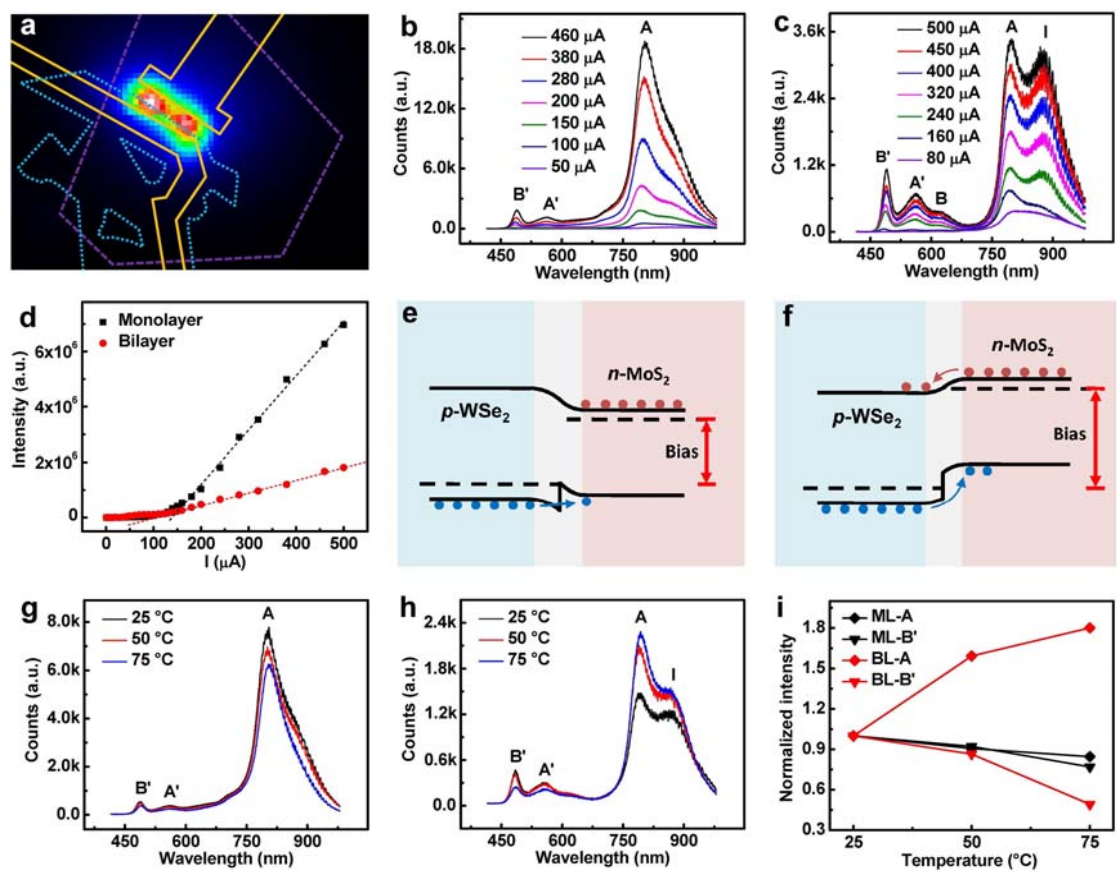

Figure 5. Electroluminescence (EL) from the WSe₂/MoS₂ heterojunction *p-n* diode. (a) The false colour EL image of the heterojunction device under an injection current of 100 µA. The purple dashed line outlines the ML-WSe₂, the blue dotted line outlines the MoS₂ and the golden solid line outlines the gold electrodes. (b) The EL spectra of a ML-WSe₂/MoS₂ heterojunction at different injection current. (c) The EL spectra of a BL-WSe₂/MoS₂ heterojunction at different injection current. (d) The EL intensity as a function of injection current for both ML- and BL- WSe₂/MoS₂ heterojunction. (e) The ideal band diagram of the WSe₂/MoS₂ heterojunction under small forward bias. The conduction band in MoS₂ is below that in WSe₂, the valence band in WSe₂ is below that in MoS₂. At small bias, holes can go cross the junction and inject into n-type region, while the electrons cannot go cross the junction. (f) The ideal band diagram of the WSe₂/MoS₂ heterojunction under large forward bias. The conduction band in MoS₂ shifts upward and is higher than that in WSe₂, and the valence band in WSe₂ is below it in MoS₂. At large bias, both electrons and holes can go cross the junction and inject into the other side of the heterojunction. (g) The EL spectra of a ML-WSe₂/MoS₂ heterojunction at different temperature ranging from 25 to 75 °C. (h) The EL spectra of a BL-WSe₂/MoS₂ heterojunction at different temperature ranging from 25 to 75 °C. The injection current is fixed at 250 µA. (i) The normalized intensities of A and B' peaks in the EL spectra of both ML- and BL-WSe₂ as a function of temperature.